# Rethinking Build vs. Buy Decisions in Enterprise Software: Navigating Trade-offs through a Structured Decision-Support Approach

Janardan Misra[1], Vikrant Kaulgud, Adam Burden, and Sanjay Podder
{janardan.misra,vikrant.kaulgud,adam.p.burden,sanjay.podder}@accenture.com

Accenture Inc.

**Build-versus-buy decisions remain a persistent challenge in enterprise software development, shaped by competing strategic, technical, cost, and risk considerations. The increasing availability of third-party solutions alongside the growing feasibility of custom development through cloud-native technologies, APIs, and low-code platforms has further amplified the complexity of these decisions. In practice, organizations often rely on fragmented expertise and informal reasoning, making it difficult to systematically analyze trade-offs or justify decisions over time. This paper presents a structured decision-support approach designed to augment build-versus-buy decision-making in such contexts. The approach is grounded in an ontology of decision factors spanning strategic considerations, application characteristics, cost and budget constraints, and risk dimensions. It combines this factor model with rule-based reasoning and reference-level matching to support decision-making even in cold-start scenarios where historical data is unavailable. The approach is implemented as a lightweight advisory artifact that enables users to evaluate relevant factors, explore trade-offs, and derive recommendations with transparent reasoning. The applicability of the approach is illustrated through a finance domain case, demonstrating how structured factor analysis can clarify decision rationale and highlight conditions under which decisions may change over time. The results suggest that making decision criteria explicit and systematically comparable can improve the quality, transparency, and auditability of build-versus-buy decisions in enterprise settings.**

# INTRODUCTION

Build-versus-buy decisions remain a persistent challenge in enterprise software development. When organizations introduce new capabilities, they must choose between adopting ready-to-use third-party solutions and building custom systems tailored to their context. This choice is rarely straightforward: it is shaped by competing considerations such as strategic differentiation, requirements fit, cost and budget constraints, time-to-market pressures, and multiple forms of technical and operational risk. Recent technology trends have further intensified this challenge. On one hand, the availability of commercial and open-source solutions has expanded rapidly; on the other, advances in cloud-native architectures, APIs, and low-code/no-code platforms have lowered the barrier to custom development. As a result, build-versus-buy decisions increasingly involve navigating complex trade-offs rather than optimizing along a single dimension.

In practice, such decisions are often made under **cold-start conditions**, where limited historical data is available and organizational knowledge is distributed across stakeholders. Consequently, decision-making frequently relies on fragmented expertise and informal reasoning, making it difficult to ensure that all relevant factors are considered systematically or that decisions can be transparently justified and revisited over time. This creates a need for structured approaches that can support decision-makers in organizing trade-offs, articulating assumptions, and reasoning about alternatives in a consistent manner.

This paper presents the design and application of a **decision-support artifact** aimed at addressing this need. The approach is grounded in an ontology of factors that captures key dimensions influencing build-versus-buy decisions, including strategic considerations, application characteristics, cost and budget constraints, and risk factors. Building on this structured representation, the artifact combines rule-based reasoning with reference-level matching to support decision-making even in the absence of historical data. The resulting system enables users to evaluate relevant factors, compare alternatives, and derive recommendations with transparent reasoning, while accommodating partial

[1] Corresponding Author

and evolving information.

The paper further demonstrates the use of this approach through a finance domain case, illustrating how structured factor analysis can clarify decision rationale and highlight conditions under which decisions may change over time. Taken together, the work contributes a practical, ontology-driven decision-support approach for build-versus-buy scenarios, along with insights into its application in enterprise settings. The paper reports a single-case application of the approach and derives lessons learned for its use under cold-start conditions.

# RELATED WORKS

The build-versus-buy decision has been studied extensively in software engineering and related domains over the past decades. Prior work has primarily focused on identifying and analyzing the factors that influence such decisions, including strategic considerations, cost, requirements fit, system reliability, and time-to-market pressures. Empirical studies based on industry surveys and interviews have consistently shown that these decisions are multi-criteria in nature and often depend on organizational context and stakeholder judgment [2], [4]. Systematic literature reviews have further consolidated these findings, highlighting that cost, performance, and reliability are among the most commonly considered criteria, while also noting the variability in decision processes across organizations [3], [7].

A significant portion of the literature has also explored formal decision models and optimization-based approaches for component sourcing and vendor selection. For example, Cortellessa et al. [1] propose an optimization framework for build-or-buy decisions in software architecture, while other studies examine component selection and sourcing strategies across in-house, COTS, OSS, and outsourced solutions [5], [6]. While these approaches provide structured formulations, they typically rely on predefined variables and assumptions, which can limit their applicability in dynamic enterprise settings where decision contexts vary widely and not all relevant factors can be specified upfront. As a result, in practice, build-versus-buy decisions continue to rely heavily on expert judgment and ad hoc reasoning, particularly in scenarios where historical data is limited or unavailable.

More recent work shas shifted attention toward **decision support in software architecture**, including approaches based on Architectural Knowledge Management (AKM) and Architecture Decision Records (ADRs). For example, recent studies have explored the use of large language models to assist in generating architectural design decisions and documenting them as ADRs [8]. These approaches demonstrate the potential of AI-assisted systems to support decision-making, but they primarily focus on generating or documenting decisions based on available architectural knowledge rather than structuring complex trade-offs across business, technical, and economic dimensions. Similarly, tools such as ArchMind have been proposed to assist architects by leveraging system context and prior knowledge to improve design decisions [9], often evaluated in controlled or semi-structured settings. Recent interest in AI-assisted architectural decision support highlights the growing importance of structured reasoning and traceability, reinforcing the need for decision-support approaches that remain effective even without extensive historical data [8], [9].

In contrast to these approaches, the focus of this paper is not on generating decisions or automating architectural reasoning, but on **supporting enterprise build-versus-buy decisions under cold-start and partial-information conditions**. The proposed approach emphasizes structuring decision-making through an ontology of factors, enabling systematic comparison of alternatives, and providing transparent reasoning that can be revisited and audited over time. This distinction is important in enterprise contexts, where decisions often involve cross-cutting trade-offs between strategy, cost, risk, and technical requirements, and where explainability and traceability are as critical as the decision outcome itself.

Overall, while prior work has established the importance of decision factors and explored various forms of decision support, there remains a gap in approaches that explicitly address **enterprise build-versus-buy trade-offs in the absence of historical data**, while also supporting structured reasoning and auditability. The work presented in this paper aims to address this gap through an ontology-driven decision-support artifact designed for practical use in such settings.

# APPLICATION SCENARIOS

This paper focuses on **build-versus-buy decision-making under cold-start conditions**, which is the most common deployment scenario in enterprise settings. In such situations, organizations must evaluate alternatives without access to a sufficiently rich repository of comparable historical decisions, and often with only partial and evolving information about requirements, constraints, and risks.

Under these conditions, decision-makers typically rely on a combination of expert judgment, informal heuristics, and fragmented documentation. While valuable, this approach makes it difficult to ensure that all relevant factors are considered systematically, that trade-offs are evaluated consistently across stakeholders, or that decisions can be transparently justified and revisited over time. These challenges are particularly pronounced for build-versus-buy decisions, which span strategic, technical, financial, and organizational dimensions.

The decision-support approach presented in this paper is explicitly designed for this cold-start context. Rather than assuming the availability of historical decision data or pre-trained models, the approach structures decision-making around an ontology of factors and supports reasoning through reference-level matching and expert rules. This enables organizations to analyze build-versus-buy trade-offs in a disciplined manner, even when prior cases are unavailable or only partially comparable.

By focusing on this deployment scenario, the paper addresses a practical and recurring challenge faced by enterprise software teams and motivates the design choices underlying the proposed decision-support artifact.

# ONTOLOGY OF FACTORS

Build-versus-buy decisions in enterprise software are rarely driven by a single dominant criterion. Instead, they emerge from interactions between strategic intent, application-specific demands, financial constraints, and multiple forms of risk. In practice, these considerations are often discussed implicitly, with different stakeholders emphasizing different aspects based on their roles and perspectives. This makes decision-making vulnerable to omission of important factors, misalignment between stakeholders, and difficulty in justifying or revisiting decisions over time.

The ontology presented in this paper organizes build-versus-buy considerations into four categories—**strategic factors, application characteristics, cost and budget, and risk factors**—to explicitly structure these decision conversations. Each category corresponds to a distinct dimension along which build-versus-buy trade-offs are commonly evaluated in enterprise contexts. Strategic factors capture long-term business concerns such as competitive differentiation, standards compliance, and intellectual property protection. Application characteristics reflect the nature of the problem being solved, including distinctiveness, complexity, non-functional requirements, and dependencies. Cost and budget factors represent the financial and resource constraints that shape feasibility and sustainability. Risk factors capture uncertainties associated with both building and buying, including technical, vendor-related, and organizational risks.

This categorization is not intended as a theoretical taxonomy, but as a **practical decision scaffold**. By grouping related factors, the ontology helps decision-makers systematically examine trade-offs that might otherwise be considered in isolation or overlooked entirely. It also supports structured dialogue across roles—for example, between architects, delivery leads, and business stakeholders—by making explicit which dimensions are being emphasized and which assumptions underpin the discussion. In cold-start situations, where historical data is unavailable, this structured representation provides a common reference point for reasoning about alternatives and documenting decision rationale.

Together, these four categories enable the advisory approach to balance completeness with usability: they are broad enough to capture the key forces shaping build-versus-buy decisions, yet sufficiently structured to support consistent analysis, explanation, and later auditability across enterprise scenarios.

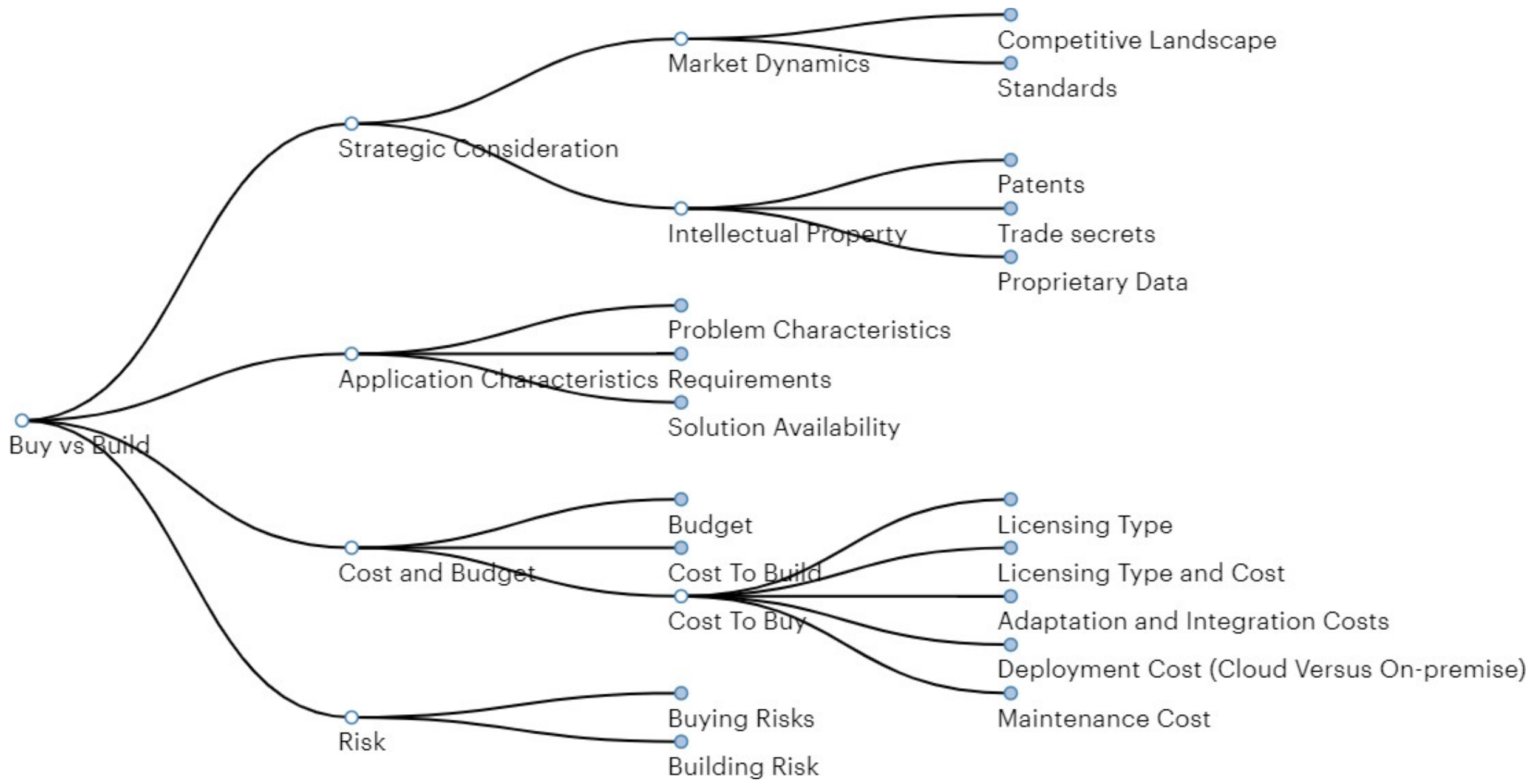

Fig. 1. Ontological Representation of Factors for Build vs Buy

Figure 1 graphically provides a high-level overview of the ontology of factors as a tree. It is designed with measurable factors across four main categories: strategic factors, application characteristics, cost and budget, and risk factors described next.

# DETAILED DESIGN OF BUILD VS BUY ADVISOR

This section describes the design of the Build vs. Buy Advisory (BBA) approach as a **decision-support artifact** intended for enterprise settings characterized by partial information and cold-start conditions. The design emphasizes transparency, usability, and support for structured reasoning rather than automation or optimization.

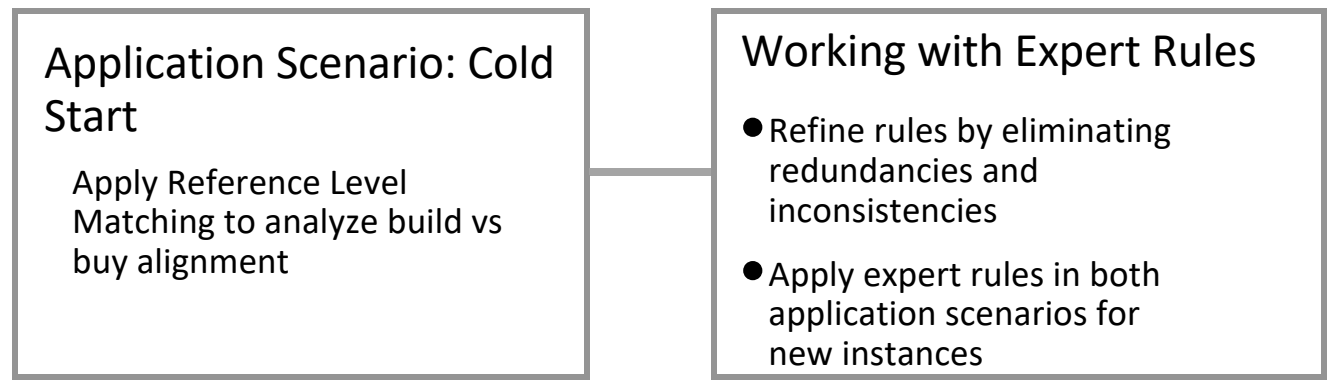

Fig. 2. High Level Process-Flow of Build vs Buy Advisor

## A. Design Overview

At a high level, the BBA approach structures build-versus-buy decision-making around three complementary elements:

1. An ontology of decision factors
2. Reference-level matching for factor-based comparison, and
3. Expert rules that encode strong, recurring organizational heuristics.

Figure 2 illustrates the overall flow of the advisory process. Users provide information about the decision scenario by specifying values for a subset of relevant factors drawn from the ontology. The advisory then analyzes these inputs to support a structured comparison between build and buy alternatives and produces recommendations accompanied by transparent reasoning.

Importantly, the design does not assume the availability of historical decision data or fully specified inputs. Instead,

it is explicitly designed to support **incremental and partial decision process**, reflecting how build-versus-buy discussions typically unfold in practice.

### B. Reference-Level Matching

Reference-level matching provides the core mechanism for structuring trade-off analysis under cold-start conditions. For each factor in the ontology, the advisory defines indicative reference values that tend to favor a build decision and corresponding values that tend to favor a buy decision when the factor is considered in isolation. For example, the presence of proprietary intellectual property may favor building, whereas strong availability of mature third-party solutions may favor buying.

When a new decision scenario is evaluated, the advisory compares the user-provided factor values against these reference levels. Rather than producing probabilistic scores or confidence measures, the system highlights **alignment patterns**, that is, which factors currently support build, which support buy, and which remain ambiguous or unspecified. This comparison allows decision-makers to see how different dimensions contribute to the emerging recommendation and where trade-offs are concentrated.

This mechanism serves two practical purposes. First, it helps ensure that relevant factors are considered systematically rather than selectively. Second, it supports reflective discussion by making explicit which assumptions drive the recommendation and which changes would be most likely to alter the decision in the future.

### C. Expert Rules as Encoded Organizational Knowledge

In addition to reference-level matching, the BBA approach supports the use of **expert rules**. These rules capture strong, experience-based heuristics that are commonly applied in enterprise build-versus-buy decisions. Rather than formal decision logic, they represent **organizational knowledge** distilled from repeated practice—for example, situations in which buying is strongly preferred due to compliance constraints, or cases in which building is unavoidable due to lack of suitable market offerings.

Expert rules are expressed as conditional patterns over factor values and are evaluated alongside reference-level matching. When applicable, they provide decisive guidance and explanatory context, reflecting well-understood organizational policies or risk thresholds. When no rule applies, the advisory falls back on factor-based comparison, ensuring that the absence of rules does not block structured analysis.

This combination allows the advisory to balance flexibility with decisiveness: expert rules encode high-confidence knowledge where it exists, while reference-level matching supports nuanced reasoning in less clear-cut situations.

### D. Supporting Explanation and Auditability

A key design goal of the BBA approach is to support explanation and auditability. For each recommendation, the advisory surfaces the factors and rules that contributed to the outcome, enabling users to understand not only *what* decision is suggested, but *why*. This is particularly important in enterprise settings where build-versus-buy decisions often need to be justified to multiple stakeholders and revisited as constraints evolve.

By structuring decisions around explicit factors, reference comparisons, and expert rules, the advisory supports traceable reasoning without requiring formal models or historical training data. The resulting design aligns with practical decision-making needs while remaining lightweight enough for real-world adoption.

## PROTOTYPE OF THE BUILD VS. BUY ADVISORY

A prototype of the Build vs. Buy Advisory (BBA) was implemented as a lightweight web-based tool to evaluate the practicality of the proposed approach and to support its use in real decision-making contexts. The prototype operationalizes the ontology, reference-level matching, and expert rules described earlier, and is designed to align with how build-versus-buy decisions are typically explored in enterprise settings.

### A. Inputs to the Advisory

The prototype accepts inputs in the form of factor values drawn from the build-versus-buy ontology. Each factor is presented to users as a question with a small set of predefined options reflecting realistic decision scenarios (e.g., levels of in-house expertise, availability of market solutions, degree of proprietary data, or security requirements). Inputs can be provided either through an online form or through a pre-filled spreadsheet, allowing flexibility in how decision data is captured.

Importantly, users are not required to provide values for all factors. Instead, they may specify only those factors that are relevant or sufficiently understood at a given point in the decision process. This design reflects the incremental and exploratory nature of enterprise decision-making, where complete information is rarely available upfront.

### B. Outputs and Recommendations

Based on the provided inputs, the prototype produces recommendations at two levels:

1. Category-level insights corresponding to the major ontology dimensions (strategic factors, application characteristics, cost and budget, and risk), and
2. An overall build-versus-buy recommendation: The outputs are qualitative and explanatory apart from an overall % score indicating faction of factors favoring Buy vs Build. They indicate which factors currently support a build decision, which support a buy decision, and where information is ambiguous or missing. This allows users to understand how different considerations contribute to the recommendation without reducing the decision to a single opaque score.

### C. Support for Partial Information

Support for partial information is a central design goal of the prototype. The advisory evaluates only the factors that have been specified by the user and does not assume completeness of input. When factors are unspecified, they are explicitly treated as neutral rather than inferred or imputed. This ensures that recommendations reflect the current state of knowledge rather than an artificial assumption of certainty.

By making gaps in information visible, the prototype encourages users to identify which additional inputs would be most valuable to refine the decision. This feature is particularly important in cold-start scenarios, where decision-making must proceed despite incomplete data.

### D. Explanation and Auditability

The prototype is designed to support explanation and auditability of build-versus-buy decisions. For each recommendation, it surfaces the specific factors and, where applicable, expert rules that influenced the outcome. Users can see how individual factor values align with build-oriented or buy-oriented reference levels and how these alignments aggregate across categories.

This explicit trace from inputs to recommendation allows decisions to be justified to stakeholders and revisited as conditions evolve. Rather than presenting recommendations as definitive outcomes, the prototype frames them as structured interpretations of current assumptions, supporting reflective discussion and long-term decision governance in enterprise environments.

## CASE STUDY: APPLYING THE BUILD VS. BUY ADVISORY IN A FINANCE DOMAIN CONTEXT

This section illustrates the use of the proposed Build vs. Buy Advisory (BBA) through a finance domain case and focuses on **how the artifact was used in practice**, rather than only on the resulting recommendation.

### A. Case Context

The case concerns a Cross-Border Payments System to be operated by a central monetary authority. The system supports regulated, high-reliability financial transactions and must meet stringent security and compliance requirements. From an organizational perspective, the system is not intended to provide competitive differentiation, but it is mission-critical and subject to long-term operational and regulatory constraints.

At the outset, both build and buy options were considered plausible. Custom development offered greater control and flexibility but implied significant development effort, specialized skills, and long-term maintenance responsibility. At the same time, several third-party solutions were available that appeared functionally adequate but raised questions around fit, cost, and long-term risk. This made the scenario representative of the build-versus-buy trade-offs commonly faced in enterprise financial systems.

### B. Use of the Advisory in the Decision Process

Factor values for the advisory were provided by the customer's stakeholders, based on their understanding of the system requirements, organizational constraints, and market options at the time of analysis. The BBA was then used as a **facilitation artifact** during the decision discussion rather than as an automated decision engine.

In this context, the primary role of the advisory was to provide a **complete and structured view of the decision space**. By organizing inputs across strategic factors, application characteristics, cost and budget, and risk dimensions, the artifact helped participants surface considerations that had not been explicitly discussed earlier. Several aspects—such as long-term maintenance implications and vendor-related risks—were examined more systematically than in prior informal discussions.

The specific instantiation of the decision scenario used in the advisory is summarized in **Table II**, which captures the selected factor values across the four ontology categories along with their corresponding build-oriented and buy-oriented reference levels. The table provides a consolidated view of how the customer's inputs map onto the decision space, making explicit the alignment of each factor with build and buy preferences. For example, factors such as solution availability, time-to-market constraints, and vendor maturity align more strongly with a buy preference, whereas aspects such as high domain criticality and stringent security requirements remain neutral or equally important for both options. This structured representation enabled stakeholders to examine the interplay between different dimensions of the decision in a consistent manner and to identify which factors were most influential in shaping the overall recommendation.

## C. Clarifying Trade-offs and Decision Rationale

While the eventual recommendation favored a buy option, the primary value of the advisory lay in clarifying the underlying trade-offs rather than merely producing the outcome. As shown in **Table II**, several factors—such as solution availability, shorter time-to-market requirements, and the presence of mature vendor support—aligned more strongly with a buy preference. At the same time, other factors, including high domain criticality and stringent security and reliability requirements, did not clearly favor either option and therefore required careful consideration rather than acting as decisive signals.

This contrast between strongly aligned and neutral factors helped stakeholders distinguish between **decision-driving considerations** and **contextual constraints**. For example, while high security requirements were initially perceived as an argument for building in-house, the structured comparison revealed that both build and buy options would need to satisfy similar security expectations, thereby reducing its role as a differentiating factor in the decision. In contrast, factors such as expected build effort, internal skill availability, and long-term maintenance burden emerged as more influential in shaping the recommendation.

The explicit mapping of factor values to reference levels, as captured in Table II, also enabled stakeholders to reason about **how the decision might evolve over time**. In particular, the analysis highlighted that improvements in internal expertise, reductions in build effort, or changes in cost structures could shift the balance toward a build decision. By making these dependencies visible, the advisory supported not only the immediate decision but also a forward-looking understanding of the conditions under which the decision would need to be revisited.

## D. Additional Usage Context

Beyond this detailed case, the advisory has also been used internally in a small number of exploratory analyses during proposal and solution-evaluation activities. In these settings, the artifact served a similar role: structuring early-stage build-versus-buy discussions under limited information and supporting transparent comparison of alternatives. While these uses are not reported as full case studies here, they reinforced the practicality of the approach in cold-start enterprise scenarios.

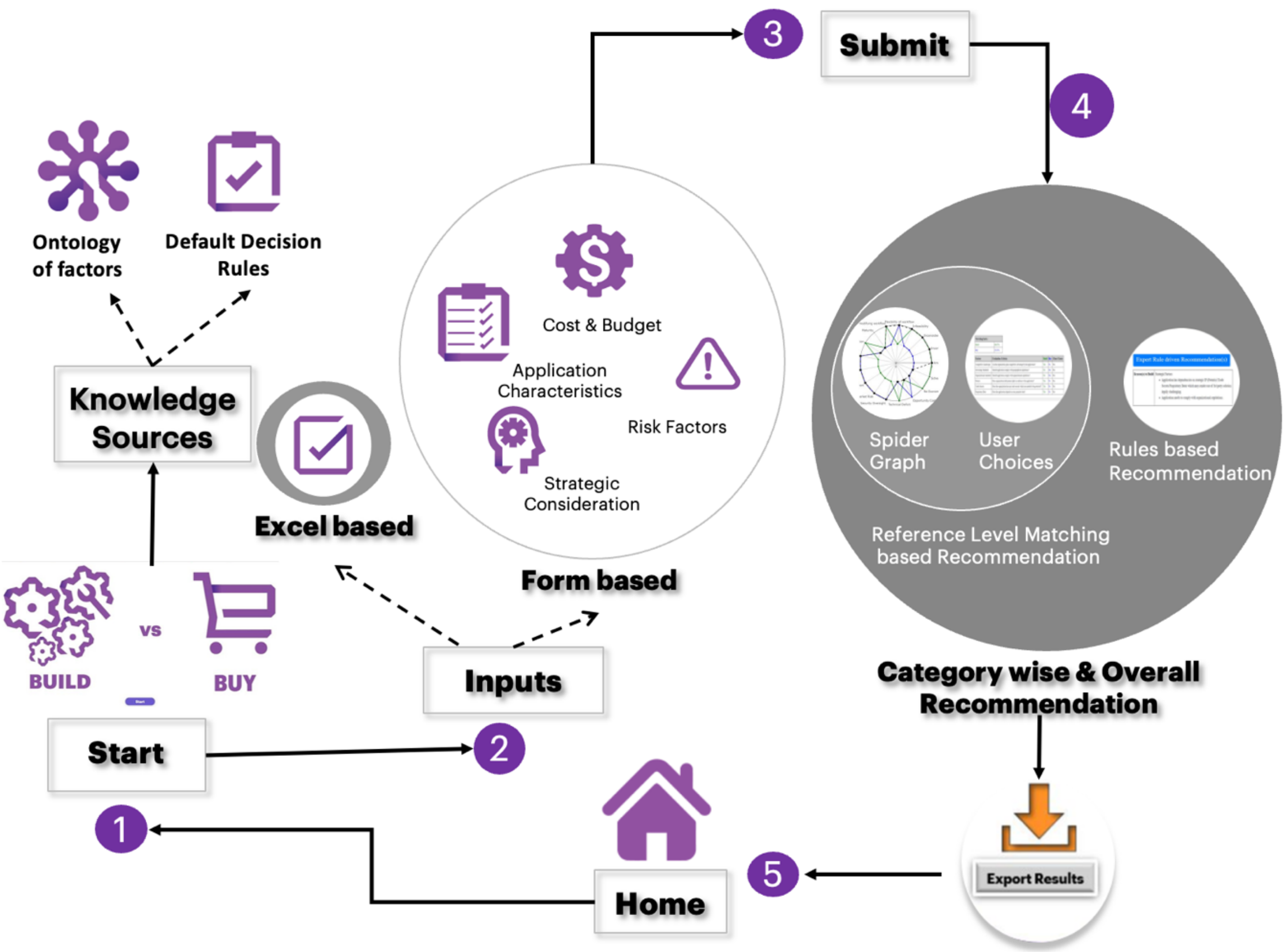

Fig. 5. UI Flow of the Build vs Buy Advisory Agent

TABLE II
CAPTURING APPLICATION SCENARIO IN TERMS OF FACTORS

| | **Factor** | **Client Scenario** | **Build Reference Level** | **Buy Reference Level** |
|---|---|---|---|---|
| **Strategic Factors** | | | | |
| $sf_1$ | Competitive advantage | No | Yes | No |
| $sf_2$ | Compliance to standards | High | Low | High |
| $sf_3$ | Proprietary IP requirement | No | Yes | No |
| **Application Characteristics** | | | | |
| $ac_1$ | Domain criticality | High | High | Low |
| $ac_2$ | Problem complexity | High | Low | High |
| $ac_3$ | Time to market | Short | Long | Short |
| $ac_4$ | Security and reliability | High | High | High |
| **Cost and Budget** | | | | |
| $cb_1$ | Build time and effort | High | Low | High |
| $cb_2$ | Licensing and deployment cost | Manageable | High | Manageable |
| $cb_3$ | Maintenance cost | High for build | Low | High |
| **Risk Factors** | | | | |
| $rf_1$ | Schedule risk | High for build | Low | High |
| $rf_2$ | Vendor maturity and support | Available | Not applicable | Available |
| $rf_3$ | Security oversight | High | High | High |

# LESSONS LEARNED

The design and application of the Build vs. Buy Advisory (BBA) across the finance domain case yielded several practical lessons relevant to enterprise software decision-making.

**Explicit factor ontologies broaden decision conversations:** Structuring the decision space through an explicit ontology helped ensure that discussions extended beyond immediately visible concerns such as cost and timelines. In the case study, strategic and risk-related factors—particularly long-term maintenance implications and vendor-related risks—were examined more systematically than in earlier informal discussions. This broadened perspective reduced the likelihood of overlooking considerations that tend to emerge only later in the system lifecycle.

**Support for partial inputs is essential in enterprise settings:** The advisory was used in a context where not all information was available upfront. Allowing users to provide only a subset of factor values enabled productive analysis without forcing premature assumptions. Making unspecified factors explicit rather than inferred helped participants identify where additional information would most improve confidence in the decision, reflecting how build-versus-buy evaluations typically evolve in practice.

**Explanation supports reflection, not just justification:** A key value of the advisory lay in its ability to explain *why* a particular direction was favored. Beyond supporting justification of the final recommendation, the explanation made it clear which factors were most influential and which changes could plausibly alter the decision in the future. In the case study, this helped stakeholders identify "switch conditions"—such as improved internal expertise or reduced build effort—that would warrant revisiting the decision at a later stage.

**A single artifact can support both decision-making and auditability:** By making assumptions, factor values, and reasoning explicit, the advisory served both as a facilitation tool during decision discussions and as a basis for documenting decision rationale. This dual role is particularly valuable in enterprise contexts, where build-versus-buy decisions often need to be revisited or justified to new stakeholders over time.

# LIMITATIONS AND THREATS TO VALIDITY

This work has several limitations that should be considered when interpreting the results.

First, the paper reports **one detailed case study**. While this case is representative of common enterprise build-versus-buy scenarios, the findings cannot be generalized without further application across additional domains and organizational contexts.

Second, the study does not include a **comparative baseline** against alternative decision-support approaches or purely informal decision-making. As a result, the paper does not claim improvements in decision quality relative to other methods, but instead focuses on demonstrating practical feasibility and structured reasoning support.

Third, the quality of recommendations depends on the **elicitation of factor values**. Inaccurate or biased inputs may affect outcomes, reflecting a broader challenge inherent in decision-support systems rather than a limitation specific to this approach.

Finally, the **generalizability of the ontology and rules** remains to be validated through broader deployment. While initial usage in exploratory analyses and proposal contexts suggests practical applicability, additional studies are needed to assess adaptability across industries, organizational scales, and regulatory environments.

Despite these limitations, the work provides a grounded starting point for structured build-versus-buy decision support under cold-start conditions and highlights directions for future validation and refinement.

# CONCLUSION AND FUTURE WORK

In this paper, we presented the design and application of a Build vs. Buy Advisory tool that can support businesses and software professionals in structuring enterprise software sourcing decisions. We developed an ontology of decision factors that consolidates considerations otherwise distributed across the literature and organizes them into strategic, application, cost, budget, and risk dimensions. The resulting tool can be used to capture such decisions systematically, support transparent reasoning, and enable future audits as assumptions and decision conditions evolve.

Future work will extend the ontology with additional factors, model relationships among them, and apply the advisory across multiple practical scenarios to identify common patterns and temporal trends in how build-versus-buy trade-offs evolve within and across industries.